\input epsf
\documentstyle[11pt]{article}

\begin{document}

\topmargin 0pt
\oddsidemargin 5mm

\setcounter{page}{1}
\begin{titlepage}
\rightline{Preprint YERPHI-1467(4)-96}
\vspace{2cm}
\begin{center}

{\bf
Double Charged Higgs Bosons Production in $e^-e^-$-Collisions}\\
\vspace{5mm}
{\large R.A.Alanakyan}\\
\vspace{5mm}
{\em Theoretical Physics Department,
Yerevan Physics Institute,
Alikhanian Brothers St.2,

 Yerevan 375036, Armenia\\}
 {E-mail: Alanak @ vx1.YERPHI.AM\\}
\end{center}

\vspace{5mm}
\centerline{{\bf{Abstract}}}
In the framework of the models with Higgs triplets, double charged Higgs
bosons production in the processes
 $e^-e^-\rightarrow\delta ^{--}_{L,R}\gamma $ are considered.
\vspace{5mm}
\vfill
\centerline{{\bf{Yerevan Physics Institute}}}
\centerline{{\bf{Yerevan 1996}}}

\end{titlepage}

 Double charged Higgs bosons arise in  theories with Higgs
sector enlarged  by triplets of Higgs bosons(see e.g. \cite{G} and ref.
therein ).Their introduction provides a natural explanation of the
smallness of the left neutrinos masses.
Double charged Higgs bosons and Majorana  neutrinos lead to some
new phenomena such as neutrinoless $\beta$
-decays, $\mu\rightarrow 3e$ decay, muonium-antimuonium conversion
 and other processes  with lepton
number violation \cite{JG, HM}.

In particular, in \cite{JG, JE} the process:
\begin{equation}
\label{A1}
e^-e^-\rightarrow \mu^-\mu^-
 \end{equation}
mediated by $\delta_{L,R}^{--}$-bosons and the  processes \cite{JE,LBN}:
\begin{equation}
\label{A2}
e^-e^-\rightarrow W^{-}_{L,R}W^{-}_{L,R}
\end{equation}
mediated by double
charged Higgs bosons or (and) heavy  Majorana neutrino
 have been considered.
High energy and high luminosity $e^-e^-$-colliders in
 particular
$e^-e^-$-version of NLC and TLC colliders have been considered in \cite{R, P}.

Here we study double charged  Higgs boson production in  the
processes
\begin{equation}
\label{A3}
 e^-e^-\rightarrow \delta^{--}_{L,R}\gamma,
\end{equation}
described by three diagrams on Fig.1.
Produced $\delta^{--}_{L,R}$-bosons  may decay into $l^-l^-$
or into $W^-_{L,R}W^-_{L,R}$-pairs if it is kinematically possible
\cite{G}.

Using formula (A5) in Appendix A for
$\delta^{--}_{L,R}$-interaction with electrons we obtain the
following gauge invariant amplitude
of the process (3):
\begin{equation}
\label{FF}
M=2eh_{ee}\bar{u}(k_1)\left( \frac{\hat{k}_4
\hat{A}}{(k_2-k_3)^2}+\frac{\hat{A}\hat{k}_4}
{(k_1-k_3)^2}+4\frac{(k_4A)}{s-m_H^2}\right) P_{L,R}u^c(k_2)
\end{equation} 
Here we neglect electron mass and use the following notations: $A_\mu$
is the polarization 4-vector of the photon, $s=(k_1+k_2)^2$, $m_H$
is the  mass of $\delta^{--}_L$ or $\delta^{--}_R$-bosons.

For differential cross section we obtain  the following result:
\begin{equation}
\label{FG}
\frac{d\sigma }{d\cos\theta }=\frac{\alpha h_{ee}^2}{s}\left(
1+\frac{2(1-\beta )}{\beta  ^2}\right) \beta ctg^2 \theta
\end{equation}

Here $\theta $ -is an angle between photon momentum $\vec{k_3}$  and  electron
momentum $\vec{k_1}$, $\beta=\left(1-\frac{m_H^2}{s}\right)$
is  the velosity of $\delta^{--}_{L,R}$-boson in the c.m.
system.

We see that our  result contains collinear
singularity at $\theta=\pm 0$, and we cut some cone  near this direction
as it has been done
 for  $e^+e^-\rightarrow Z^0\gamma$  process (see \cite{B}  and
 references therein).

The cross section of the process  (3) as well as cross section of the
process $e^+e^-\rightarrow Z^0\gamma$ contain also
infrared singularity near reaction threshold.
 
 Number of events $\delta^{--}_{L,R}\gamma $ per year
 ($\sigma$$L$) is shown on Fig.2
at $\sqrt{s}=0.5,1$ TeV  and luminosity  $L=10^{41}sm^{-2}$.
We use cut $|\cos\theta| < 0.9$ and 0.95.

Thus we see that consideration of the process (3) may provide new
restriction on the $h_{ee}$ and $m_H$  in addition to the restriction
from nonobservation of above mentioned low energy
processes with lepton number violation, anomalous muons magnetic
moment and Bhabba scattering \cite{G}-\cite{HM}.

Let us compare the cross section of the process (1) and (2) with
the cross  section of the studied process (3).

The process $e^-e^-\rightarrow W^{-}_{R}W^{-}_{R}$
  will be  kinematically forbidden for  large masses
of $ W_R^{\pm}$-bosons  $(2m_{W_R}>\sqrt{s})$.

The process $e^-e^-\rightarrow \delta_L^{--*}\rightarrow\ W^{-}_{L}W^{-}_{L}$
 may be suppresed by smallness of the vertex
$W_L^-W_L^-\delta ^{++}_{L}$
.For instance, in left-right models this vertex is suppressed by
factor $\frac{v_L}{k_L} $
which is small for preserving true
relation between $W_L^{\pm}$,$Z^0$-bosons masses and Weinberg's angle.

The cross section of the process (1) is of order
$h_{ee}^2h^2_{\mu\mu}$   whereas
the cross section of the studied process is of order $h^2_{ee}$,  so
the cross section of the process (3)  at    small $h_{\mu\mu}$ and far
from resonanse  (i.e. far from range $\sqrt{s}=m_H$)  may
dominate over reaction   (1).

It must be noted, that all our results are also applicable for a more general
case where Yucawa couplings of the left triplet and right triplet with
left- and right-handed leptons are different.

\indent
\newpage
\setcounter{equation}{0}
\appendix{{\bf Appendix A}}

\renewcommand{\theequation}{A.\arabic{equation}}
\indent
In the left-right symmetric model  the interaction
 of left and right triplets with
($Y=2$) of
Higgs bosons:
\begin{equation}
\label{AD}
\Delta  _{L,R}= \left(\begin{array}{ll}
 \delta  ^+_{L,R}/\sqrt{2}&
\delta  ^{++}_{L,R}\\
\delta  ^0_{L,R}& - \delta^+_{L,R}/\sqrt{2}
\end{array}\right)
\end{equation}
with left- and right-handed lepton fields $\psi^T_{L,R}=(\nu
_{L,R}^T,e^T_{L,R}) $       are  described by
lagrangian:
\begin{equation}
\label{DD}
{\cal L}=  ih _{ij}\left(\psi^ T_{iL}
C \tau _2\Delta _L\psi_{jL}
+\psi^{T}_{iR} C\tau _2 \Delta _R \psi_{jR} \right)+ h.c.
\end{equation}

Here $i$,$j=e,\mu,\tau $-are generations indices, $C$ is the  charge
conjugation matrix, $\tau_2$ is the Pauli matrix.
After symmetry breaking Majorana masses of the heavy approximately
right handed neutrinos are expressed through the Yucawa couplings
$h$ and neutral component of right triplet vacuum expectation
$v_R$
in the following  way:

\begin{equation}
\label{DE}
m_{N}=\sqrt{2}hv_{R}
\end{equation}
Also, large right triplet vacuum expectation ($v_L\ll k_L$,$k_R\ll v_R$,
$k_L,k_R$- are vacuum expectations of the left and right doublets,
$v_L$-vacuum expectation of the left triplet) 

 provide mass of the
$W_{R}^{\pm}$-bosons:
\begin{equation}
\label{DF}
m_{W_{R}}=\frac{1}{2}gv_R
\end{equation}
whereas doublet vacuum expectation is responsible for mass of
 $W_{L}^{\pm}$-bosons.

So, as seen from (A3),(A4) in left-right symmetric models Yucawa couplings $h$
are expressed through the $m_{W_R}$ and $m_N$.

From (A1), (A2)
$e^-e^-\rightarrow \delta^{--}_{L,R}$  transition is given
by amplitude:
\begin{equation}
\label{A5}
{\cal{M}}=2h_{ee}\bar{u}(k_1)P_{L,R}u^c(k_2)
\end{equation}
where $u^c=C\bar{u}^T$ ,$k_1$, $k_2$-are momenta of the two electrons.

In general, mass matrix of $\delta^{--}_{L,R}$ bosons is nondiagonal,
however in
the limit  $v_L\ll  v_R$ mixing between  $\delta_L^{--}$   and
$\delta_R^{--}$   is negligible.

\newpage

\centerline{\bf Figures captions:}
Fig.1 Diagramms corresponding to the processes $e^-e^-\rightarrow
\delta ^{--}_{L,R}\gamma $           .

Fig.2 Number of events $\delta^{--}_{L,R}\gamma $ per year
(at $L=10^{41}sm^{-2}$) produced in reaction (3)
as a function of $m_{H}$ at $h_{ee}=10^{-2}$.Solid lines 1,2 correspond
to the energies $\sqrt{s}=0.5,1$ TeV and cut $|\cos\theta| < 0.9$.

Dotted curves 3,4 correspond to the energies
 $\sqrt{s}=0.5,1 $ TeV and cut $|\cos\theta| < 0.95 $.
\begin{titlepage}
\begin{figure}[htb]
\epsfxsize=15cm
\epsfysize=8.5cm
\mbox{\hskip -1.0in}\epsfbox{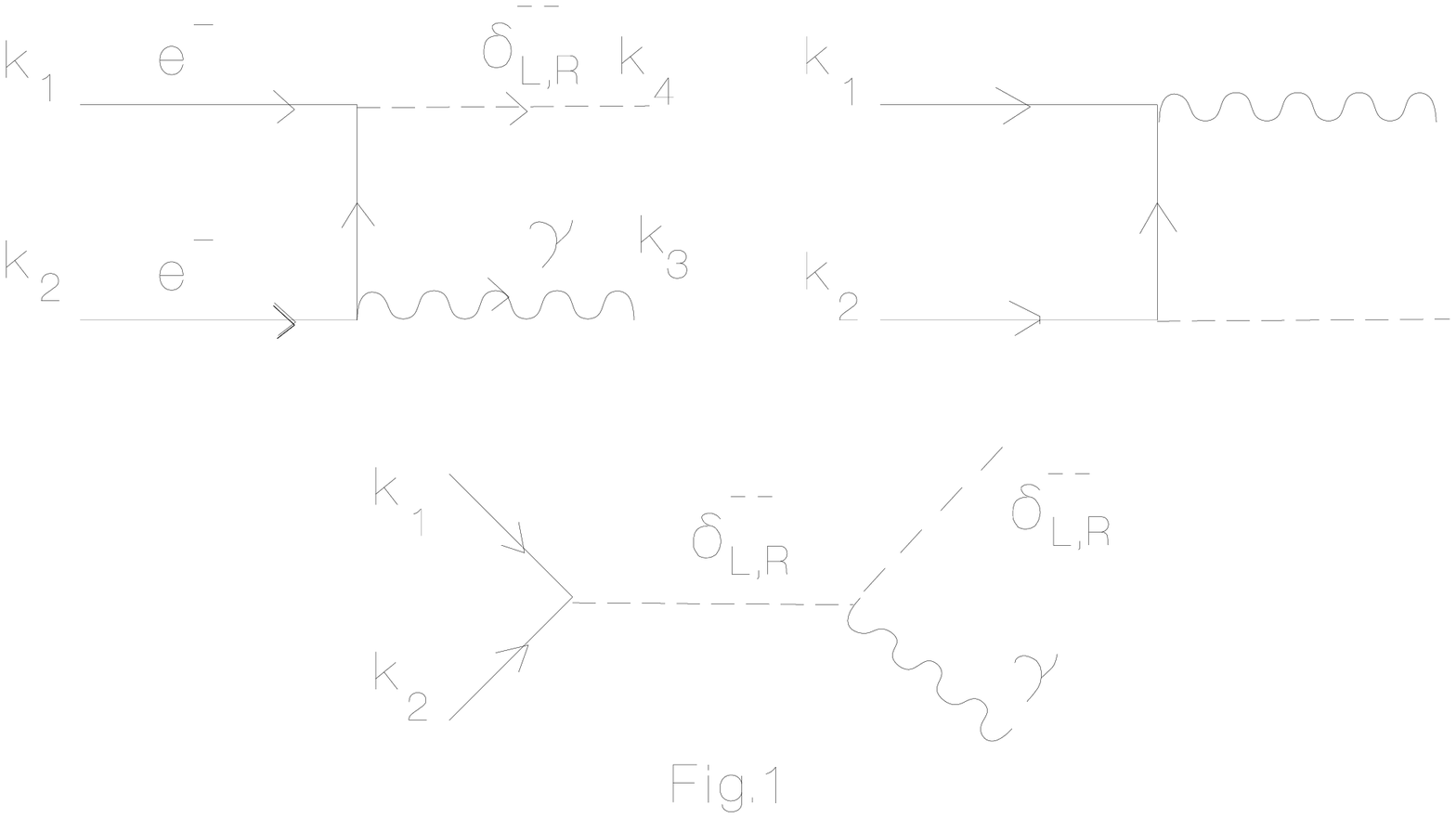}
\end{figure}
\begin{figure}[htb]
\epsfxsize=15cm
\epsfysize=8.5cm
\mbox{\hskip -1.0in}\epsfbox{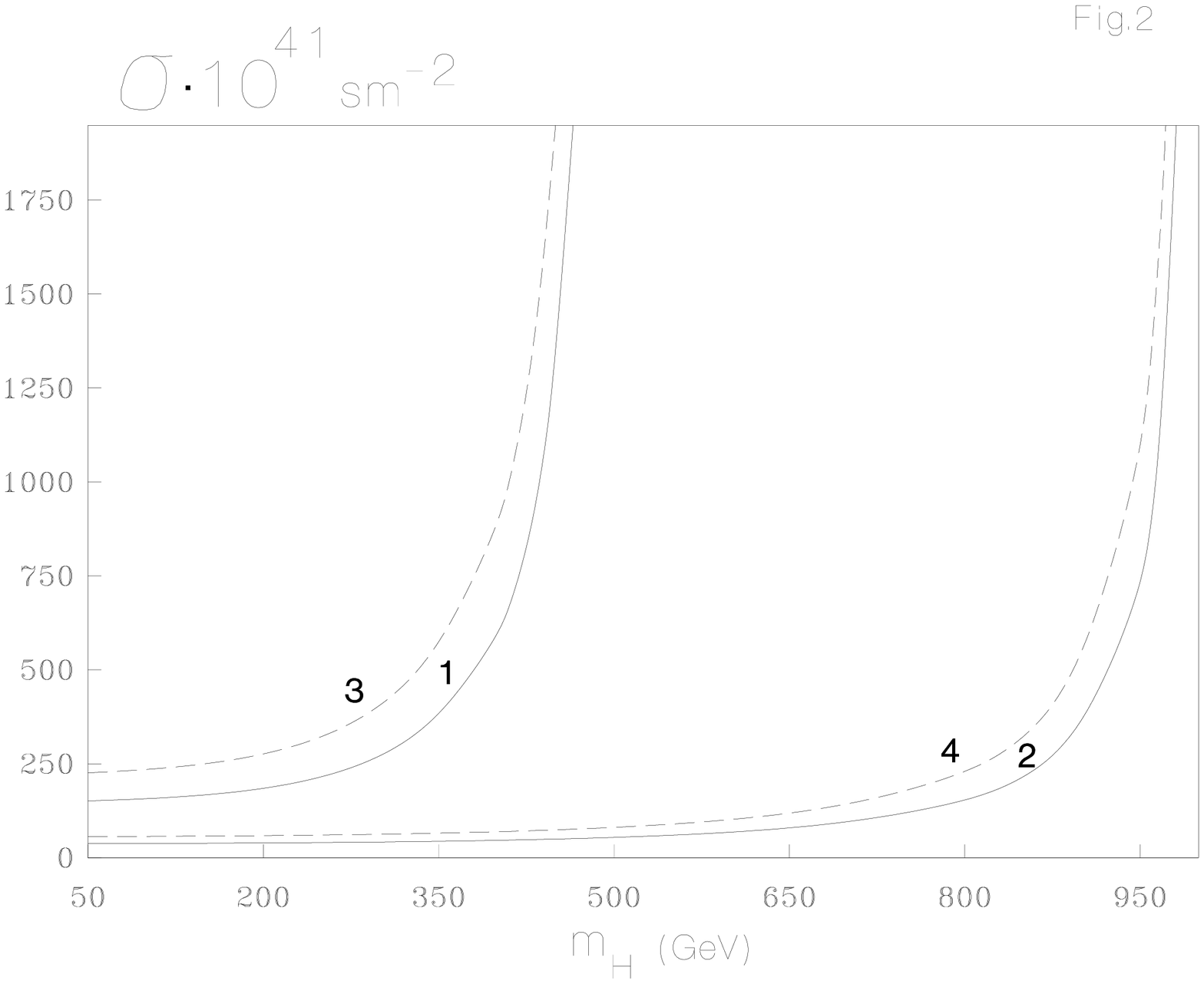}
\end{figure}
\end{titlepage}
\end{document}